\documentclass[final]{svjour3}
\pdfoutput=1
\usepackage{gensymb}
\usepackage{graphicx}
\usepackage{rotating}
\usepackage{amssymb}
\usepackage{mathptmx}
\usepackage{amsmath}
\usepackage[numbers]{natbib}
\usepackage[table,xcdraw]{xcolor}
\usepackage{bm}
\makeatletter
\journalname{Journal of Low Temperature Physics}

\bibpunct{}{}{,}{s}{}{,}

\begin{document}

\newcommand{\hdblarrow}{H\makebox[0.9ex][l]{$\downdownarrows$}-}
\title{Optical characterization of the Keck Array and BICEP3 CMB Polarimeters from 2016 to 2019}

\author{T.~St.~Germaine$^1$ \and P.~A.~R.~Ade$^2$ \and Z.~Ahmed$^{3,4}$ \and M.~Amiri$^5$ \and D.~Barkats$^1$ \and R.~Basu Thakur$^6$ \and C.~A.~Bischoff$^7$ \and J.~J.~Bock$^{6,8}$ \and H.~Boenish$^1$ \and E.~Bullock$^9$ \and V.~Buza$^1$ \and J.~Cheshire$^9$ \and J.~Connors$^1$ \and J.~Cornelison$^1$ \and M.~Crumrine$^9$ \and A.~Cukierman$^{3,4}$ \and M.~Dierickx$^1$ \and L.~Duband$^{10}$ \and S.~Fatigoni$^5$ \and J.~P.~Filippini$^{11,12}$ \and S.~Fliescher$^9$ \and J.~A.~Grayson$^4$ \and G.~Hall$^{1,9}$ \and M.~Halpern$^5$ \and S.~Harrison$^1$ \and S.~R.~Hildebrandt$^8$ \and G.~C.~Hilton$^{13}$ \and H.~Hui$^6$ \and K.~D.~Irwin$^{3,4}$ \and J.~Kang$^4$ \and K.~S.~Karkare$^{1,14}$ \and E.~Karpel$^4$ \and S.~Kefeli$^6$ \and S.~A.~Kernasovskiy$^4$ \and J.~M.~Kovac$^1$ \and C.~L.~Kuo$^4$ \and K.~Lau$^9$ \and E.~M.~Leitch$^{14}$ \and K.~G.~Megerian$^8$ \and L.~Moncelsi$^6$ \and T.~Namikawa$^{15}$ \and C.~B.~Netterfield$^{16,17}$ \and H.~T.~Nguyen$^{6,8}$ \and R.~O'Brient$^{6,8}$ \and R.~W.~ Ogburn~IV$^4$ \and S.~Palladino$^7$ \and C.~Pryke$^9$ \and B.~Racine$^1$ \and C.~D.~Reintsema$^{13}$ \and S.~Richter$^1$ \and A.~Schillaci$^6$ \and R.~Schwarz$^9$ \and C.~D.~Sheehy$^{18}$ \and A.~Soliman$^6$ \and B.~Steinbach$^6$ \and R.~V.~Sudiwala$^2$ \and K.~L.~Thompson$^4$ \and J.~E.~Tolan$^4$ \and C.~Tucker$^2$ \and A.~D.~Turner$^8$ \and C.~Umilt\`{a}$^7$ \and A.~G.~Vieregg$^{14}$ \and A.~Wandui$^6$ \and A.~C.~Weber$^8$ \and D.~V.~Wiebe$^5$ \and J.~Willmert$^9$ \and C.~L.~Wong$^1$ \and W.~L.~K.~Wu$^{14}$ \and E.~Yang$^{3,4}$ \and K.~W.~Yoon$^4$ \and E.~Young$^4$ \and C.~Yu$^4$ \and C.~Zhang$^6$}

\institute{\email{stgermaine@g.harvard.edu}
\\$^1$Harvard-Smithsonian Center for Astrophysics, Cambridge, Massachusetts 02138, USA
\\$^2$School of Physics and Astronomy, Cardiff University, Cardiff, CF24 3AA, United Kingdom
\\$^3$Kavli Institute for Particle Astrophysics and Cosmology, SLAC National Accelerator Laboratory, 2575 Sand Hill Rd, Menlo Park, California 94025, USA
\\$^4$Department of Physics, Stanford University, Stanford, California 94305, USA
\\$^5$Department of Physics and Astronomy, University of British Columbia, Vancouver, British Columbia, V6T 1Z1, Canada
\\$^6$Department of Physics, California Institute of Technology, Pasadena, California 91125, USA
\\$^7$Department of Physics, University of Cincinnati, Cincinnati, Ohio 45221, USA
\\$^8$Jet Propulsion Laboratory, Pasadena, California 91109, USA
\\$^9$Minnesota Institute for Astrophysics, University of Minnesota, Minneapolis, 55455, USA
\\$^{10}$Service des Basses Temp\'{e}ratures, Commissariat \`{a} lEnergie Atomique, 38054 Grenoble, France
\\$^{11}$Department of Physics, University of Illinois at Urbana-Champaign, Urbana, Illinois 61801, USA
\\$^{12}$Department of Astronomy, University of Illinois at Urbana-Champaign, Urbana, Illinois 61801, USA
\\$^{13}$National Institute of Standards and Technology, Boulder, Colorado 80305, USA
\\$^{14}$Kavli Institute for Cosmological Physics, University of Chicago, Chicago, IL 60637, USA
\\$^{15}$Department of Applied Mathematics and Theoretical Physics, University of Cambridge, Wilberforce Road, Cambridge CB3 0WA, UK
\\$^{16}$Department of Physics, University of Toronto, Toronto, Ontario, M5S 1A7, Canada
\\$^{17}$Canadian Institute for Advanced Research, Toronto, Ontario, M5G 1Z8, Canada
\\$^{18}$Physics Department, Brookhaven National Laboratory, Upton, NY 11973}

\maketitle

\begin{abstract}

The BICEP/\textit{Keck} experiment (BK) is a series of small-aperture refracting telescopes observing degree-scale Cosmic Microwave Background (CMB) polarization from the South Pole in search of a primordial $B$-mode signature. This $B$-mode signal arises from primordial gravitational waves interacting with the CMB, and has amplitude parametrized by the tensor-to-scalar ratio $r$. Since 2016, BICEP3 and the Keck Array have been observing with 4800 total antenna-coupled transition-edge sensor detectors, with frequency bands spanning 95, 150, 220, and 270 GHz. Here we present the optical performance of these receivers from 2016 to 2019, including far-field beams measured in situ with an improved chopped thermal source and instrument spectral response measured with a field-deployable Fourier Transform Spectrometer. As a pair differencing experiment, an important systematic that must be controlled is the differential beam response between the co-located, orthogonally polarized detectors. We generate per-detector far-field beam maps and the corresponding differential beam mismatch that is used to estimate the temperature-to-polarization leakage in our CMB maps and to give feedback on detector and optics fabrication. The differential beam parameters presented here were estimated using improved low-level beam map analysis techniques, including efficient removal of non-Gaussian noise as well as improved spatial masking. These techniques help minimize systematic uncertainty in the beam analysis, with the goal of constraining the bias on $r$ induced by temperature-to-polarization leakage to be subdominant to the statistical uncertainty. This is essential as we progress to higher detector counts in the next generation of CMB experiments.

\keywords{Cosmic Microwave Background, Inflation, Polarization, Transition-Edge Sensor, Keck Array, BICEP3}

\end{abstract}

\section{Introduction}

The Cosmic Microwave Background (CMB) contains a wealth of information on the origin and evolution of the universe, and has been observed by many ground-based and satellite instruments over the past few decades.  These observations have helped constrain the parameters of the standard $\Lambda$CDM model used to describe cosmic evolution.  There exist significant problems with this model, namely the horizon, flatness, and monopole problems.  These can be solved with an extension to the $\Lambda$CDM model -- a period of brief and exponential inflation early in the universe.  Inflation generates both scalar and tensor perturbations to the metric, which manifest as density waves and gravitational waves, respectively.  These primordial gravitational waves interact with the surface of last scattering, imprinting both a pure-gradient $E$-mode pattern and a pure-curl $B$-mode pattern in the polarization of the CMB.  The $B$-mode pattern is a unique prediction of inflation and peaks at degree angular scales.   The amplitude of this signal is parametrized by the tensor-to-scalar ratio $r$.  A non-zero measurement of $r$ (after accounting for other $B$-mode signals from Galactic foregrounds and gravitational lensing) would be strong evidence of the existence of this period of inflation.

The BICEP/\textit{Keck} series of experiments has been targeting this signal from an observing site at the South Pole since 2006.  The compact on-axis refracting telescope design allows for sensitivity to CMB polarization at degree angular scales ($\ell \sim $100).  The latest published results include all data taken by BICEP2 and \textit{Keck Array} up through the observing year 2015 (BK15), with data at observing frequencies of 150 GHz, 95 GHz, and 220 GHz.  When combined with \textit{Planck} temperature measurements and other external data, BK15 yields a constraint of $r < 0.06$ at 95\% C.L. [1].

The \textit{Keck Array} has been observing since 2011, and has a telescope design very similar to BICEP2.  It features a 264 mm aperture diameter, a field of view of 15$\degree$, and $f/2.4$ optics including lenses made from high-density polyethylene (HDPE).  The detectors are transition-edge sensor (TES) bolometers [2] with time-domain multiplexed readout enabled by superconducting quantum interference devices (SQUIDs).  The light couples to each detector pair via two co-located, orthogonally polarized slot antenna arrays.  Each \textit{Keck} receiver holds 272 (95 GHz) or 496 (150/220/270 GHz) optically-coupled detectors.

BICEP3 has been observing since 2015, and features a larger aperture (520 mm), wider field of view (27$\degree$), and fast optics ($f/1.7$) including lenses made from alumina ceramic [3].  The detectors are antenna-coupled TES bolometers, similar to \textit{Keck}.  The SQUID-based readout is also similar, except the MUX chips are installed behind the detectors in a detector module, allowing for modularity and higher packing efficiency.  The BICEP3 focal plane unit (FPU) holds 2400 optically-coupled detectors.

To measure polarization, we take the difference between each co-located, orthogonally polarized detector pair.  Any mismatch in the beam pattern leaks CMB temperature into polarization.  Since this may lead to a false detection of $B$ modes, it is imperative that we measure our beams and precisely characterize the difference pattern in each pair.  For this reason, we spend 1--2 months at the end of every Austral summer measuring our beams \textit{in situ}.  In analysis we apply a technique called deprojection [4], which filters out any coupling of CMB temperature to a second-order expansion of the beam profile.  For the BK15 dataset, when the measured temperature to polarization ($T \rightarrow P$) leakage is added to simulations, the bias on the value of $r$ recovered from the multicomponent likelihood analysis is $\Delta r = 0.0027 \pm 0.0019$ [5].

In these proceedings we present beam measurements taken on the \textit{Keck Array} and BICEP3 from 2016 to 2019, and progress toward improving the systematics in the beam analysis, which directly impacts our ability to estimate the final bias on $r$ from $T \rightarrow P$ leakage.  In Section 2 we show bandpass spectra measurements, and in Section 3 we show far-field beam map measurements, including 2D Gaussian fits to the beams as well as the high-fidelity array-averaged beam maps.  We conclude and look ahead to future work in Section 4.

\section{Bandpass Measurements}

In order to detect the $B$-mode signal from primordial gravitational waves, we must separate this signal from potential galactic foregrounds such as polarized thermal dust and synchrotron emission.  Since each of these sources has a different frequency dependence, we can constrain these foregrounds by observing with multiple frequencies.  Therefore we need to know the spectral response in each of our observing bands in order to constrain the foregrounds in our multicomponent likelihood analysis.

\begin{figure}[t]
\begin{center}
\includegraphics[width=0.7\linewidth, keepaspectratio]{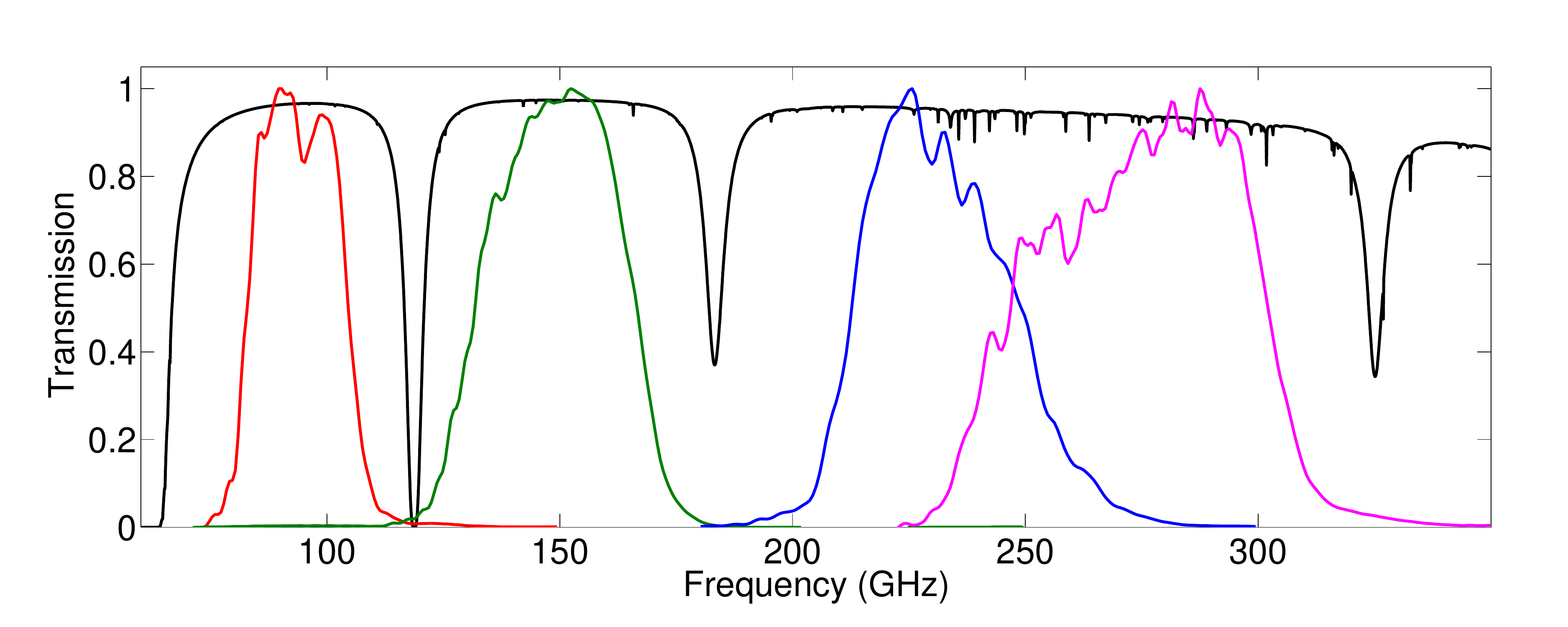}
\caption{Measured array-averaged, peak-normalized spectra for all observing bands from 2016 to 2019.  From left to right: BICEP3 (band center 94.2 GHz, band width 26.7 GHz), \textit{Keck} 150 (149.0 GHz, 43.4 GHz), \textit{Keck} 220 (231.2 GHz, 51.2 GHz), \textit{Keck} 270 (275.4 GHz, 69.8 GHz).  These spectra are response to a beam-filling source described by a Rayleigh-Jeans spectrum. The typical winter South Pole atmospheric transmission is overlaid in black, calculated using \texttt{am} [8].  (Color figure online.)}
\label{fig:bandpass}
\end{center}
\end{figure}

To measure the bandpass of each detector, we use a Martin-Puplett Fourier Transform Spectrometer (FTS) that can be mounted directly onto any receiver \textit{in situ} [6].  This spectrometer has spectral resolution of 0.5 GHz and has a thermal source that is differential between 77 K and ambient temperature ($\sim 250$ K at South Pole).  The detectors are biased on the aluminum TES superconducting transition in order to avoid saturation.  We measure spectra for every working detector, and coadd them to form array-averaged spectra for each observing frequency.  Table~\ref{table1} shows the observing bands in each year from 2016 to 2019, and the corresponding measured array-averaged spectra are shown in Fig.~\ref{fig:bandpass}.  The spectra show a 25--28\% band width in each band, which is defined as $\Delta \nu = (\int S(\nu) \, d \nu)^2 / \int S(\nu)^2 \, d \nu$ for frequency $\nu$ and spectral response $S(\nu)$.


\begin{table}[]
\centering
\begin{tabular}{c|cccc}
                  & \textbf{2016}                  & \textbf{2017}                  & \textbf{2018}                  & \textbf{2019}                  \\ \hline
\textbf{BICEP3}   & {\color[HTML]{FE0000} 95 GHz}  & {\color[HTML]{FE0000} 95 GHz}  & {\color[HTML]{FE0000} 95 GHz}  & {\color[HTML]{FE0000} 95 GHz}  \\
\textbf{Keck Rx0} & {\color[HTML]{3166FF} 220 GHz} & {\color[HTML]{3166FF} 220 GHz} & {\color[HTML]{3166FF} 220 GHz} & {\color[HTML]{3166FF} 220 GHz} \\
\textbf{Keck Rx1} & {\color[HTML]{3166FF} 220 GHz} & {\color[HTML]{3166FF} 220 GHz} & {\color[HTML]{3166FF} 220 GHz} & {\color[HTML]{009901} 150 GHz} \\
\textbf{Keck Rx2} & {\color[HTML]{3166FF} 220 GHz} & {\color[HTML]{3166FF} 220 GHz} & {\color[HTML]{3166FF} 220 GHz} & {\color[HTML]{3166FF} 220 GHz} \\
\textbf{Keck Rx3} & {\color[HTML]{3166FF} 220 GHz} & {\color[HTML]{3166FF} 220 GHz} & {\color[HTML]{3166FF} 220 GHz} & {\color[HTML]{3166FF} 220 GHz} \\
\textbf{Keck Rx4} & {\color[HTML]{009901} 150 GHz} & {\color[HTML]{6200C9} 270 GHz} & {\color[HTML]{6200C9} 270 GHz} & {\color[HTML]{6200C9} 270 GHz}
\end{tabular}
\caption{Nominal observing band centers for BICEP3 and \textit{Keck Array} from 2016 to 2019.  Note that BICEP3 has 2400 detectors and each \textit{Keck} receiver listed here has 496 detectors. The 150 GHz receiver in 2019 was retrofitted with microwave multiplexing readout and SLAC Microresonator Radio Frequency warm electronics [7]. (Color online.)}
\label{table1}
\end{table}

\section{Far-Field Beam Map Measurements}

The small apertures of the BICEP/\textit{Keck} telescopes give a far-field distance $2 D^2 / \lambda$ that is less than 200 m for all of our observing frequencies.  Since BICEP3 and \textit{Keck Array} are housed in separate buildings $\sim$200 m apart, we observe a chopped thermal source placed on the opposite building to map our detectors.  The source is chopped at $\sim$16 Hz between ambient temperature microwave absorber ($\sim$250 K) and the sky at zenith ($\sim$10 K).  This procedure has been discussed in many  publications; for a complete description, see the BK15 Beams paper [5].

\subsection{Coordinate System and 2D Gaussian Fits}

For our beam maps we define an instrument-fixed spherical coordinate system that is independent of the instrument orientation with respect to the sky.  A pixel $P$ containing two orthogonally polarized detectors has a defined location $(r,\theta)$ with respect to the boresight $B$ (the telecentric axis), where $r$ is the radial distance from $B$ and $\theta$ is the counterclockwise angle looking out from the telescope towards the sky from the $\theta$ = 0$\degree$ ray.  The $\theta$ = 0$\degree$ ray is fixed to the instrument: for \textit{Keck} we choose this to be the line that divides Tiles 1 and 2, and for BICEP3 we choose it to be the line that runs along Tiles 11, 10, and 9 (for a diagram of the \textit{Keck} and BICEP3 FPUs, see Fig. 2 of [5] and Fig. 3 of [9], respectively).  

From our raw beam map detector timestreams, we demodulate at the chop rate to isolate the signal from the chopped source, then bin into maps with 0.1$\degree$ square pixels.  The demodulation routine has been improved in this dataset, drastically reducing the non-Gaussian noise in the beam maps.  We then fit each beam to a 2D elliptical Gaussian with six free parameters:

\begin{equation}
	B(\bm{x}) = \frac{1}{\mathrm{\Omega}}e^{-\frac{1}{2}(\bm{x}-\bm{\mu})^{T}\Sigma^{-1}(\bm{x}-\bm{\mu})},
\end{equation}
where $\bm{x} = (x',y')$ is the beam map coordinate, $\bm{\mu} = (x_0,y_0)$ is the beam center, $\mathrm{\Omega}$ is the normalization, and $\Sigma$ is the covariance matrix, parametrized as:

\begin{equation}
    \Sigma = \quad \begin{pmatrix} 
    \sigma^2 (1 + p) & c\sigma^2 \\
    c\sigma^2 & \sigma^2(1-p) 
    \end{pmatrix}.
\end{equation}
$\sigma$ is the beamwidth, and $p$ and $c$ are plus and cross ellipticy, respectively.   A Gaussian with a major axis oriented along the $x_0$ or $y_0$ axes has $+p$ or $-p$ ellipticity, respectively, and one with a major axis oriented diagonally has $\pm c$ ellipticity.  Differential parameter estimates are defined to be that parameter for the A detector minus that for the B detector, for example $dp = p_A - p_B$.

\begin{table}[]
\centering
\resizebox{\textwidth}{!}{%
\begin{tabular}{|c|c|c|c|c|}
\hline
Parameter & Rx4 2016 (150) & Rx3 2019 (220) & Rx4 2019 (270) & BICEP3 2019 (95) \\ \hline
$\sigma (\degree)$ & 0.201 $\pm$ 0.002 $\pm$ 0.004 & 0.139 $\pm$ 0.002 $\pm$ 0.003 & 0.120 $\pm$ 0.003 $\pm$ 0.003 & 0.166 $\pm$ 0.004 $\pm$ 0.002 \\
$p$ & 0.004 $\pm$ 0.019 $\pm$ 0.029 & 0.000 $\pm$ 0.017 $\pm$ 0.037 & 0.005 $\pm$ 0.023 $\pm$ 0.033 & -0.023 $\pm$ 0.026 $\pm$ 0.018 \\
$c$ & 0.003 $\pm$ 0.014 $\pm$ 0.031 & 0.004 $\pm$ 0.031 $\pm$ 0.039 & 0.006 $\pm$ 0.041 $\pm$ 0.041 & -0.026 $\pm$ 0.031 $\pm$ 0.017 \\
$d\sigma (\degree)$ & 0.000 $\pm$ 0.001 $\pm$ 0.001 & 0.000 $\pm$ 0.000 $\pm$ 0.000 & 0.000 $\pm$ 0.001 $\pm$ 0.000 & 0.000 $\pm$ 0.001 $\pm$ 0.000 \\
$dp$ & -0.020 $\pm$ 0.005 $\pm$ 0.003 & -0.018 $\pm$ 0.004 $\pm$ 0.002 & -0.016 $\pm$ 0.007 $\pm$ 0.006 & 0.004 $\pm$ 0.012 $\pm$ 0.002 \\
$dc$ & -0.003 $\pm$ 0.002 $\pm$ 0.003 & 0.003 $\pm$ 0.004 $\pm$ 0.003 & -0.006 $\pm$ 0.011 $\pm$ 0.006 & -0.001 $\pm$ 0.004 $\pm$ 0.002 \\
$dx (')$ & 0.18 $\pm$ 0.40 $\pm$ 0.12 & -0.53 $\pm$ 0.18 $\pm$ 0.04 & -0.77 $\pm$ 0.13 $\pm$ 0.04 & -0.05 $\pm$ 0.14 $\pm$ 0.05 \\
$dy (')$ & -0.12 $\pm$ 0.33 $\pm$ 0.11 & 0.42 $\pm$ 0.11 $\pm$ 0.04 & 0.18 $\pm$ 0.25 $\pm$ 0.04 & 0.02 $\pm$ 0.16 $\pm$ 0.06 \\ \hline
\end{tabular}%
}
\caption{Beam parameter summary statistics for a representative receiver-year for each observing frequency from 2016 to 2019.  Values listed as FPU Median $\pm$ FPU Scatter $\pm$ Measurement Uncertainty.  The variation of one unchanged receiver from year to year is smaller than the measurement uncertainty.  The beamwidth has been corrected for the non-negligible size of the chopped source aperture.}
\label{table2}
\end{table}

Each detector has at least $\sim$10 different measurements taken each year.  For each detector and pair, we take the median over all these measurements to be the best estimate, and half the width of the central 68\% of the distribution of those measurements as the measurement uncertainty.  Table ~\ref{table2} shows a summary of the measured beam parameters for a representative receiver-year for each observing frequency between 2016 and 2019, presented as FPU Median $\pm$ FPU Scatter $\pm$ Measurement Uncertainty.  Comparing to previous published results (e.g. [5]), we see the beamwidths are systematically smaller in the values shown in Table ~\ref{table2}.  This is due to an improvement in the analysis pipeline to reduce beam smearing due to binning and interpolation effects.  Reducing beam smear also changes the shape of the beam window functions shown in Fig.~\ref{fig3} which directly impacts our primary CMB analysis.

Most of the power in the difference beams is contained in a second-order expansion of the beam profile, which couples to the CMB temperature and its derivatives.  This can be filtered out in a process called deprojection, where a best-fit template of the \textit{Planck} temperature map and its derivatives is scaled and removed from the pair difference data (see [4] for a full description).  The differential parameters discussed above roughly correspond to the modes that are deprojected. 

\subsection{Composite and Array-Averaged Maps}

\begin{figure}[t]
\begin{center}
\includegraphics[width=0.7\linewidth, keepaspectratio]{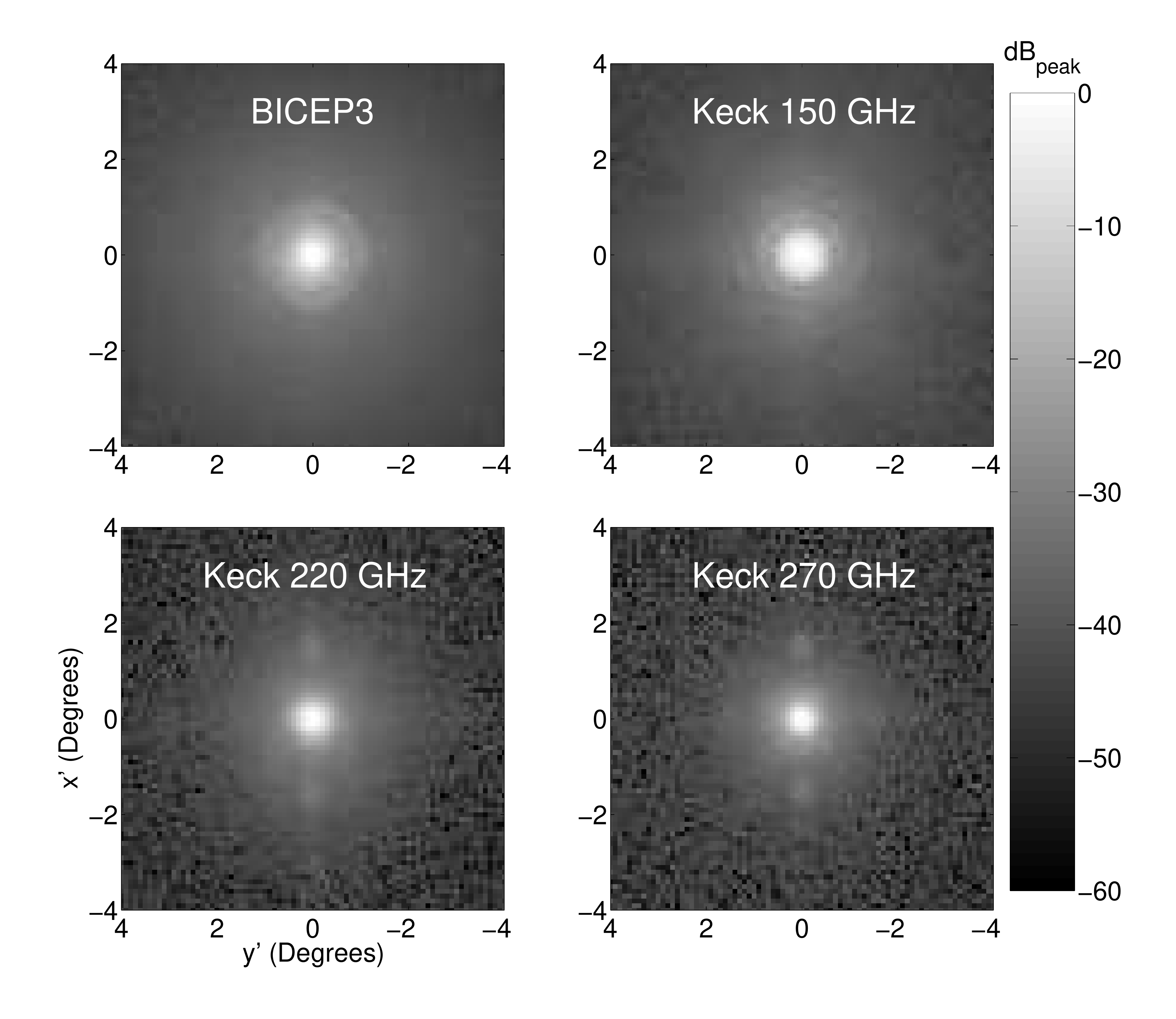}
\caption{Array-averaged beam maps for the four observing frequencies between 2016 and 2019.  Note that BICEP3 (95 GHz) has twice the aperture size of \textit{Keck Array} receivers.  Airy ring structure is clearly visible, as well as cross-talk beams from neighboring detectors in our time-domain multiplexing scheme.}
\label{fig2}
\end{center}
\end{figure}

To obtain full coverage beam maps of each detector, we coadd individual ``component" beam maps taken at multiple boresight rotation angles to obtain per-detector composite maps.  Only component maps that pass cuts are included; cuts remove beams with poor fits, beams that do not strike the mirror that redirects to the source, and detectors that were not properly biased for that run.  Azimuth-fixed contaminated signal is masked out of each component map before being coadded.  The masking routine was improved in the analysis of this dataset, further reducing contamination from the ground and rejecting all rays from the detector that do not hit the redirecting mirror.

\begin{figure}[t]
\begin{center}
\includegraphics[width=0.7\linewidth, keepaspectratio]{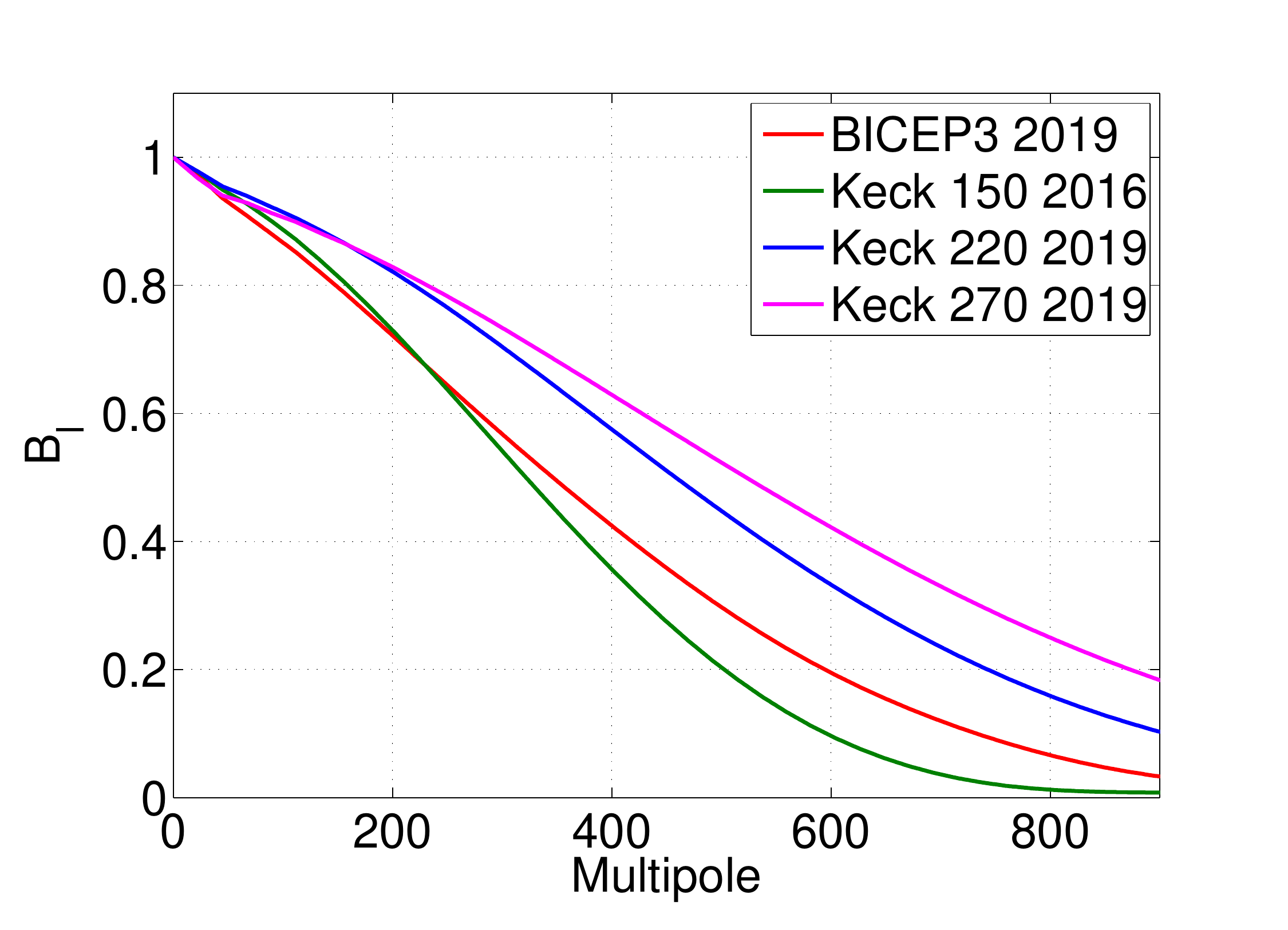}
\caption{Circularly symmetric beam window functions for each observing frequency from 2016 to 2019.  These have been corrected to account for the non-negligible size of the diameter of the chopped thermal source. (Color figure online.)}
\label{fig3}
\end{center}
\end{figure}

Once we have composites for every working detector, we integral normalize and coadd them together to form an array-averaged beam map for each observing frequency.  Fig.~\ref{fig2} shows these maps for each frequency from 2016 to 2019.  We then take the average of each of these maps in radial bins, and apply a Fourier transform to obtain the beam window function $B(\ell)$, shown in Fig.\ref{fig3}.  These are used to smooth the \textit{Planck} input maps used in our standard simulation framework.

\section{Conclusions}

We have presented optical characterization data of BICEP3 and \textit{Keck Array} receivers taken over four austral summers from 2016 to 2019, including bandpass spectra and far-field beam performance.  All spectra, including the new 270 GHz FPU, demonstrate a 25--28\% band width.  Differential beam parameters for BICEP3, taken from 2D Gaussian fits, are similar to or lower than those from \textit{Keck Array} receivers.  Composite beam maps were generated, and will be used in ``beam map simulations" to estimate the $T \rightarrow P$ leakage present in BICEP3 and \textit{Keck Array} data after deprojecting the leading order difference modes.  Analysis of beam map simulations from the BK15 dataset resulted in a bias on $r$ of $\Delta r = 0.0027 \pm 0.0019$, which we expect will improve in the next dataset after reducing the systematics in the beam maps themselves. 

While this paper has addressed analysis techniques toward minimizing beam map systematics, there has also been effort in detector design to minimize systematics due to anomalous interactions with the corrugated frame [10]. This effort toward improved detector design will continue in parallel with the analysis effort. 

\begin{acknowledgements}
The BICEP/Keck projects have been made possible through a series of grants from the National Science Foundation including 0742818, 0742592, 1044978, 1110087, 1145172, 1145143, 1145248, 1639040, 1638957, 1638978, 1638970, and 1726917, by the Gordon and Betty Moore Foundation, by the Keck Foundation, and by the grant 55802 from John Templeton Foundation. The development of antenna-coupled detector technology was supported by the JPL Research and Technology Development Fund and NASA Grants 06-ARPA206-0040, 10-SAT10-0017, 12-SAT12-0031, 14-SAT14-0009 and 16-SAT16-0002. The development and testing of focal planes were supported by the Gordon and Betty Moore Foundation at Caltech. Readout electronics were supported by a Canada Foundation for Innovation grant to UBC. The computations in this paper were run on the Odyssey cluster supported by the FAS Science Division Research Computing Group at Harvard University. The analysis effort at Stanford and SLAC is partially supported by the U.S. DoE Office of Science. We thank the staff of the U.S. Antarctic Program and in particular the South Pole Station without whose help this research would not have been possible. We thank all those who have contributed past efforts to the BICEP/Keck series of experiments, including the Bicep1 team. Tireless administrative support was provided by Kathy Deniston, Sheri Stoll, Irene Coyle, Donna Hernandez, Dana Volponi, and Julie Shih.
\end{acknowledgements}

\pagebreak

\end{document}